\documentclass{optica-article}

\journal{opticajournal} 

\articletype{Research Article}

\usepackage{mathtools}
\DeclarePairedDelimiter\ket{\lvert}{\rangle}

\usepackage{lineno}

\begin{document}

\title{Maximizing Purity and Heralding Efficiency of Type-I Down-Converted Photons Using Beam Focal Parameters}

\author{Andrew Rockovich,\authormark{1,*} Shu'an Wang,\authormark{1,2} and Daniel J. Gauthier\authormark{1}}

\address{\authormark{1}Department of Physics, The Ohio State University, Columbus, Ohio 43210, USA\\
\authormark{2}Department of Physics, Tsinghua University, Beijing, China}

\email{\authormark{*}rockovich.6@osu.edu}

\date{\today}

\begin{abstract*}
We demonstrate theoretically that correlated bi-photons can be generated with high heralding efficiency and high spectral purity for non-collinear Type-I spontaneous parametric down-conversion.  In Type-I down-conversion, the generated photons have the same linear polarization that is perpendicular to the linear pump polarization.  Previously, it was thought that high efficiency and purity could not be obtained for this configuration.  We show that the non-collinear geometry gives an additional degree-of-freedom that allows for simultaneous optimization of these source metrics. We predict near-unity ($\approx0.97$) heralding efficiency and single-photon spectral purity by adjusting the beam focal parameters, which can be obtained over a wide range of pump, signal, and idler wavelengths without requiring special crystal dispersion characteristics. As an example, we predict a heralding efficiency of 0.97, a single-photon purity of 0.97, and a pair production rate of 0.50 pairs/(s$~$mW$~$THz) using a 400-$\mu$m-long $\beta$-barium borate crystal pumped by a 355-nm-wavelength pulsed laser with a bandwidth of 8-THz. Our work offers a simple and universal approach for producing high-quality quantum photonic states for a wide variety of quantum information science applications. 
\end{abstract*}

\section{Introduction}

Photonic quantum states are ideally suited for long-distance applications such as ground-to-satellite \cite{Ren} or metropolitan \cite{Shen} communication networks. Moreover, spontaneous parametric down-conversion (SPDC) is one of the most widely used approaches for photon pair generation \cite{Couteau}.

SPDC is the nonlinear optical process whereby one incoming photon spontaneously splits into a pair of photons with strong correlations in the polarization, spatial, and temporal modes due to energy and momentum conservation of the nonlinear optical process. Using an appropriate experimental configuration, these correlations can give rise to entangled polarization \cite{Kwiat} or time-energy \cite{Steinberg} states of the photon. Entangled photons are used in diverse applications, including quantum communication \cite{Piveteau}, quantum cryptography \cite{Boaron}, quantum metrology \cite{Higgins}, etc. 

One SPDC metric that is important for many applications is the heralding efficiency $\eta$, which is the probability of measuring a partnered ``heralded'' photon upon measuring the other photon. A recently proposed approach to achieve $\eta \sim 99\%$ involves cavity-enhanced SPDC \cite{MK}. Platforms on which researchers have experimentally maximized $\eta$ include periodically poled lithium niobate \cite{Pomarico}, collinear bulk optics \cite{Ramelow}, and highly nondegenerate periodically-poled potassium titanyl phosphate crystal \cite{Kaneda}.

\begin{figure}[t!]
    \centering
    \includegraphics[width = .47\textwidth]{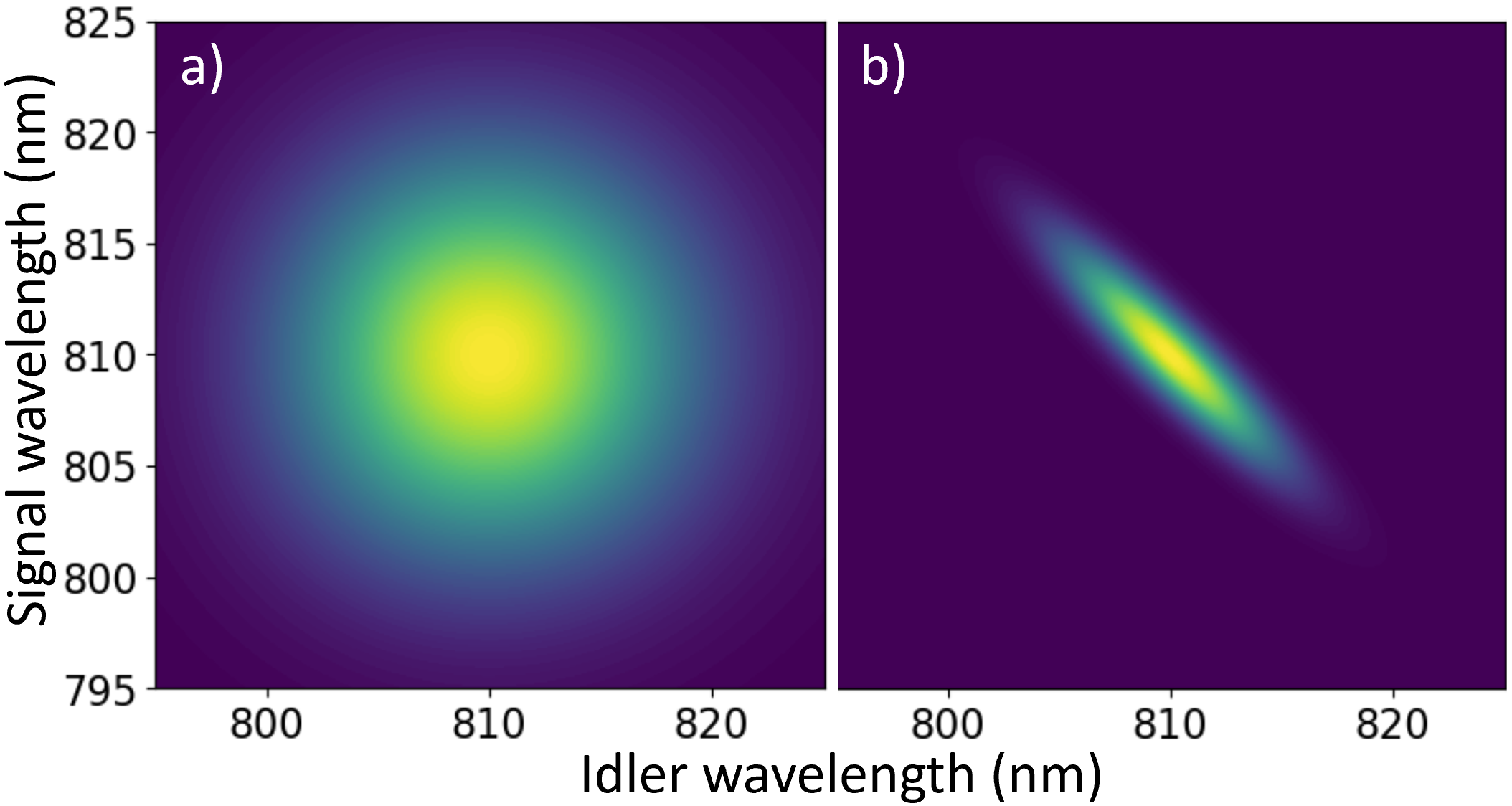}
    \caption{\label{fig:JSA} Joint spectral amplitudes (JSA) of two SPDC configurations, one with a) no frequency correlation and hence unit purity, and b) one with strong frequency correlation and low purity. The scale is from a high (bright green) to a low (dark purple) likelihood that the two photons exist at their respective wavelengths. For a monochromatic pump as considered in \cite{Guilbert}, the JSA would be infinitely narrow along the -45$^\circ$ line and the purity approaches zero.}
\end{figure}

Other applications of SPDC, such as entanglement swapping \cite{Basset} and quantum teleportation \cite{XM}, require a high degree of spectral purity $\mathcal{P}$ for photon interference \cite{Fedrizzi,Mosley}.  Here, ``purity'' refers to the degree of frequency-uncorrelation between the generated photons. A high-purity state is illustrated in Fig.~\ref{fig:JSA}a. 

High $\mathcal{P}$ is difficult to achieve using SPDC due to correlations resulting from the conservation of momentum and energy. For example, one of the generated photons can have a lower energy (momentum) as long as the other has higher energy (momentum) so that they add to the pump photon's energy (momentum). This case is shown in Fig.~\ref{fig:JSA}b, where down-converted light is highly anti-correlated in signal and idler wavelength space. 

Increasing $\mathcal{P}$ requires using a pulsed source with finite spectral width, which broadens the joint spectral amplitude (JSA) along the diagonal direction, visualized in the transition from Fig.~\ref{fig:JSA}b to Fig.~\ref{fig:JSA}a. Then, post-generation spectral filtering methods are usually used to remove the correlations, but this results in diminished $\eta$ and down-converted pair production rate $\mathcal{R}$. Therefore, developing a convenient method to achieve a high $\mathcal{P}$ without sacrificing the generation rate is desirable. 

An additional metric used to characterize these sources is the symmetrized fidelity $\mathcal{F}$. Fidelity, in general, is the overlap between a received state and the state that was expected \cite{Jozsa}. The symmetrized fidelity accounts for the fidelity of both the signal and idler beams and therefore scales with $\eta$ and $\mathcal{P}$ \cite{Meyer-Scott}.

The primary purpose of this paper is to demonstrate theoretically that high $\eta$ and $\mathcal{P}$ can be achieved simultaneously for Type-I SPDC based on laser-pumped bulk nonlinear optical crystals. We achieve our results by adjusting the focal parameters of the pump beam and the down-converted collection optics, which effects the transverse phase mismatch and hence controls the JSA.

Because we rely on transverse phase mismatch, our approach only works for non-collinear SPDC setups, where the signal and idler modes are along different axes. This is a common method for separating signal and idler modes in bulk-crystal setups. Importantly, our approach can be applied for a wide range of pump and down-converted wavelengths in contrast to techniques that only work for a limited range of wavelengths (\textit{e.g.}, group velocity matching, crystal dispersion engineering, etc.). This work adds to the quantum toolbox for sources of entangled light. 

In the next section, we discuss relevant background material such as down-converted pair source requirements and contrast our work with previous methods. In Sec.~\ref{Sec:FOM}, we discuss the various figures-of-merit that determine the quality of our proposed source. We then introduce the mechanism that allows us to tune down-conversion properties, and we show its efficacy through simulation in Sec.~\ref{Sec:TuneProp}. We discuss adjusting other physical properties of the setup and their effects on the system in Sec.~\ref{Sec:Generalize}. Section~\ref{Sec:Conclusion} summarizes our findings.

\section{Background} \label{sec:Background}

In this section, we give additional background information to set the stage for our innovation. First, we define $\eta$ as the probability of observing a photon upon observing its partner. Typically, attaining a higher $\eta$ is beneficial for entangled-photon experiments because the protocol fails if only one of the photons is received, \textit{e.g.}, quantum communication \cite{Piveteau}. Other applications require a minimum $\eta$, such as the loophole-free tests of Bell’s inequality ($\eta>66\%$ \cite{Shalm, Giustina}) and three-party quantum communication ($\eta>60\%$ \cite{Boström,Wojcik}). 

Current methods of producing pure single-photons using an SPDC source can be categorized in two ways. One approach is to engineer the crystal and optics such that the down-converted photons are pure by design. In this vein, many groups rely on engineering the crystal dispersion to achieve phase matching using an auxiliary variable to achieve group velocity matching (GVM). For example, spectral correlations can be reduced by choosing an appropriate crystal length and pump spectral bandwidth \cite{Grice}. Although this method can yield high purity ($\mathcal{P} \sim 1$), the crystal must be engineered for a specific pump wavelength, meaning there is little to no spectral flexibility \cite{Mosley}.

Engineered material dispersion methods include introducing pulse-front-tilt on the pump beam using diffraction gratings \cite{Torres}. Other groups have used apodized and chirped quasi-phase-matched periodically poled crystals \cite{Imeshev,Arbore,Nasr,Fejer}, or optimized aperiodically poled crystals \cite{Fejer,Dosseva}, etc. In general, these methods tend to be difficult or cumbersome. Systems that use some of these methods are available commercially. However, the price is invariably much higher than that of bulk BBO with some common optics, making our approach more practical for many applications.

The other approach to achieve $\mathcal{P} \sim 1$ is to use heavy spectral filtering of the down-converted pairs \cite{Pan,lu}. This is a simpler approach, but it can reduce $\mathcal{P}$ because it introduces spectrally-dependent loss that affects individual photons independently \cite{Dosseva}. Heavily filtering also reduces $\mathcal{R}$. Typical experiments that achieve $\mathcal{P} \sim 1$ using this method report count rates on the order of $\mathcal{R} \sim 10 $ pairs/(s mW) \cite{lu}, similar to our predictions in Sec.~\ref{Sec:TuneProp}. To counteract this loss, it is possible to multiplex heavily filtered sources, but this is also a difficult approach \cite{Fumihiro}. The theoretical limits on $\mathcal{P}$ and $\eta$ using only spectral filtering in a Type-II collinear single-mode waveguide are discussed by Meyer-Scott \textit{et al.} \cite{Meyer-Scott}.  They predict $\eta \sim 20\%$ as $\mathcal{P} \rightarrow 1$, and $\eta$ and $\mathcal{P}$ intersect at about $\sim 60\%$, far below what we achieve here. It follows that our approach also obtains higher $\mathcal{F}$ as well. 

In contrast, we use a simple method that resolves the difficulties of using SPDC for producing pure single-photons while maintaining high $\eta$. Our approach does not require engineering the crystal's dispersion, so it is applicable over a broad range of pump wavelengths. We use a hybrid technique: we use a non-collinear geometry, loosely spectrally filter the down-converted light, and optimize the transverse mode beam waists. Changing the beam waists is a degree-of-freedom that allows us to maximize $\mathcal{R}$ while also maximizing $\mathcal{P}$ and $\eta$.

The relationship between $\eta$, $\mathcal{P}$, and $\mathcal{R}$ has been studied previously by Bennink \cite{Bennink}, Mosley \textit{et al.} \cite{Mosley}, and Meyer-Scott \textit{et al.} \cite{Meyer-Scott} who quantified the trade-off between $\mathcal{P}$ and $\eta$.  They only consider the case of collinear SPDC, where the signal and idler modes share the same axis. Achieving unit $\mathcal{P}$ for this geometry requires that the pump photon's inverse group velocity lies between that of the signal and idler's inverse group velocities, which is only possible for Type-II SPDC (perpendicularly polarized signal and idler beams) and a narrow range of pump, signal, and idler wavelengths. 

The work by Vicent \textit{et al.} \cite{Vicent} is similar to our study in that they consider Type-I SPDC with non-collinear Gaussian beams.  They predict high $\mathcal{P}$ and $\mathcal{R}$ by adjusting the transverse beam waists, but they do not attempt to optimize $\eta$.  For their best parameters, they obtain $\eta = 0.53$. As discussed below, we obtain near unit $\mathcal{P}$ and $\eta$ but a value of $\mathcal{R}$ that is about two orders of magnitude smaller compared to their result. Although we sacrifice $\mathcal{R}$, our improves on the existing literature for experiments that require high $\mathcal{P}$ and $\eta$.

The underlying technique we use is the same as in \cite{Bennink, Shi, Vicent}\textemdash we attempt to make the joint spectral amplitude Gaussian in signal and idler frequency space (or elliptical with major and minor axes aligned to the signal and idler frequency axes). This is done by writing the joint spectral intensity (modulus-square of the JSA) in the form
\begin{equation}
\label{Eq:JSIgeneral}
    S(\Omega_s,\Omega_i)=\mathcal{N}\textrm{exp}\left({-\delta_s\Omega_s^2-\delta_i\Omega_i^2-\delta_{si}\Omega_s\Omega_i}\right),
\end{equation}
where $\Omega_j$ are the frequency detunings of the signal ($s$) and idler ($i$) photons from their phase-matched values. The spectral correlations are removed when $\delta_{si}\xrightarrow{}0$. We demonstrate the efficacy of this method in Sec.~\ref{Sec:TuneProp}.

Below, we first establish expressions for $\mathcal{R}$, $\eta$, and $\mathcal{P}$ and maximize $\eta$ or $\mathcal{P}$, or each simultaneously. We determine the optimal values for the pump beam waist and the additional parameters such as the post-generation frequency filter width, the pump spectral bandwidth ($B_p$), and crystal length $L$ and cut angle (the angle between the crystal axis and the direction of pump propagation). We find that $\mathcal{P}$ and $\eta$ are tunable up to $\mathcal{P} \sim 1$ or $\eta \sim 1$ by adjusting the down-converted collection mode beam waists for the case when the pump beam waist sets the characteristic spatial scale, defined precisely below. Finally, we discuss the trade-offs between $\eta$, $\mathcal{P}$, and $\mathcal{R}$. 

\section{SPDC Source Figures of Merit} \label{Sec:FOM}

In SPDC, a pump ($p$) photon is converted to a signal ($s$) and an idler ($i$) photon via a second-order nonlinear optical process described by the nonlinear polarization\cite{boyd},
\begin{equation}
    \vec{P} = \epsilon_0\chi^{(1)}\vec{E}+\epsilon_0\chi^{(2)}\vec{E}^2+... \label{Eq:polarizatoin}
\end{equation}
where $\chi^{(2)}$ is the second-order susceptibility tensor responsible for SPDC as well as several other nonlinear optical phenomena such as second harmonic generation (SHG). The process of splitting photons via SPDC obeys energy conservation (a ``parametric'' process), where $\omega_p=\omega_s+\omega_i$). For peak conversion efficiency, the photons obey momentum conservation ($\vec{k_p}=\vec{k_s}+\vec{k_i}$). Here, $\omega_j$ represents the angular frequency ($j$ = $p$, $s$, or $i$). Likewise, the wave vector magnitude is
\begin{equation}
\label{eq:wavevector}
    k_j = \frac{n_j(\omega_j)\omega_j}{c},
\end{equation}
where the indices of refraction $n_j$ depend on frequency (chromatic dispersion) and polarization (birefringence) \cite{born_Optics}. 

In the example below, we consider using $\beta$-barium borate (BBO), which is a negative uniaxial crystal. We assume Type-I phase-matching in which the pump is extraordinarily polarized and the signal and idler are ordinarily polarized. Projecting the three modes onto ordinary and extraordinary waves, we can represent the nonzero elements of the second-order polarization in terms of the effective second-order nonlinear susceptibility, $d_{eff}$, which is comprised of elements of the nonlinear susceptibility $2d_{il}=\chi^{(2)}_{ijk}$ using contracted notation. The terms in $d_{eff}$ are elements of the $d_{il}$ tensor that match the interaction type. For a Type-I interaction in BBO, $d_{eff}= d_{11}\cos{3\phi}\cos{\theta} - d_{31} \sin{\theta}$ \cite{boyd}. Here, $\phi$ is the angle from the $x$-axis ($=0$ because pump propagation is along $z$ and there is rotational symmetry about $z$, so we look at the $x=0=yz$ plane), and $\theta$ is the angle from the direction of pump propagation relative to the crystal axis (the cut angle). In our simulations, we first determine the cut angle that produces collinear SPDC and adjust the crystal axis slightly to achieve the desired non-collinear angles of the signal and idler beams. Further explanation on our choice of cut angle is given in Sec.~\ref{Sec:Generalize}.

The two-photon quantum state generated by the SPDC process is given by \cite{Dosseva}
\begin{equation}
    \ket*{\psi} = \int d\omega_s \int d\omega_i \Phi(\omega_s, \omega_i)\ket*{\omega_s}_s\ket*{\omega_i}_i, \label{Eq:psi}
\end{equation}
where $\ket*{\omega_s}_s$ and $\ket*{\omega_i}_i$ are the created signal and idler photon states at frequencies $\omega_s$ and $\omega_i$, respectively. The joint spectral amplitude is given by $\Phi(\omega_s, \omega_i)=g(\omega_s+\omega_i)f(\omega_s, \omega_i)$, where $g(\omega_p)$ [$f(\omega_s, \omega_i)$] is the pump [joint signal and idler] frequency distribution(s). The joint spectral intensity is given by $S(\omega_s, \omega_i) = |\Phi(\omega_s, \omega_i)|^2$.

\subsection{Absolute Pair Production Rate}

Our work builds on the previous work of Ling \textit{et al.} \cite{Ling} and Guilbert \textit{et al.} \cite{Guilbert}, who considered the Type-I SPDC process pumped by a monochromatic laser, and the work of Valencia \textit{et al.} \cite{Shi}, who consider a pulsed laser source with pulse-front tilt. The studies using monochromatic pump light adjusted the pump beam and collection beam waists to obtain $\eta \sim 1$, but monochromatic pumping gives rise to low-purity states with frequency anti-correlations such as that shown in Fig.~\ref{fig:JSA}b. Valencia \textit{et. al} obtain $\mathcal{P} \sim 1$ by adjusting the pump pulse width and wavefront tilt, but they did not attempt to optimize $\eta$ or $\mathcal{R}$. Here, we combine both theoretical approaches to answer whether all three metrics can be optimized simultaneously.

To begin, we introduce Hermite-Gauss mode functions given by
\begin{multline}
\label{Eq:HGmode}
     U_{n,m}(x,y,z)=\alpha^{(n,m)}G_n\left(\frac{\sqrt{2}x}{W(z)}\right)G_m\left(\frac{\sqrt{2}y}{W(z)}\right)\\
     \times \textrm{exp}\left(-ikz-ik\frac{x^2+y^2}{2R(z)}+i(m+n+1)\zeta\right).
\end{multline}
where  
\begin{equation}
\label{Eq:alpha}
\alpha^{(n,m)} = \sqrt{\frac{2}{2^{n+m}n!m!\pi W_0^2}},
\end{equation}
\begin{equation}
\label{Eq:Hermite_polynomial}
G_n(u)=H_n(u) \textrm{exp}\left(-\frac{u^2}{2}\right),
\end{equation}
where $R(z)$ is the radius of curvature of the beam's wavefront, $\zeta$ is the Guoy phase, $H_n(u)$ is a Hermite polynomial, and $W(z)$ is the $z$-dependent beam radius in the nonlinear crystal, defined as the $1/e$ field radius. We consider all modes to lie in the $yz$-plane, so while $x$ is the same for all modes, $y$ and $z$ are different for each ($z$ being that mode's propagation direction). The minimum beam radius $W_0$ (the `beam waist') is related to the Rayleigh range through the relation $z_r=\pi W_0^2/\lambda$, where $\lambda$ is the vacuum wavelength. We assume that the beam waists for all three modes are at the center of the crystal so that the waist for other positions is given by $W(z)=W_0\sqrt{1+(z/z_r)^2}$.

We focus on the case where $z_r \gg L$, appropriate for thin crystals often used when pumping with short-pulse laser light that we need to obtain high $\mathcal{P}$. 
Under this assumption, we can take $k(x^2+y^2)/R(z) \sim 0$ and the Guoy phase $\zeta \sim 0$, which greatly simplifies evaluating the integral in Eq.~(\ref{Eq:psi}). Working in the thin-crystal regime also allows us to consider the beam radii to be at their minima throughout the length of the crystal, so that $W_j(z)=W_{0,j}$ for the $j$-mode (where $j=s\textrm{, }i\textrm{, or }p$).

\begin{figure}[ht]
    \centering
    \includegraphics[width = .47\textwidth]{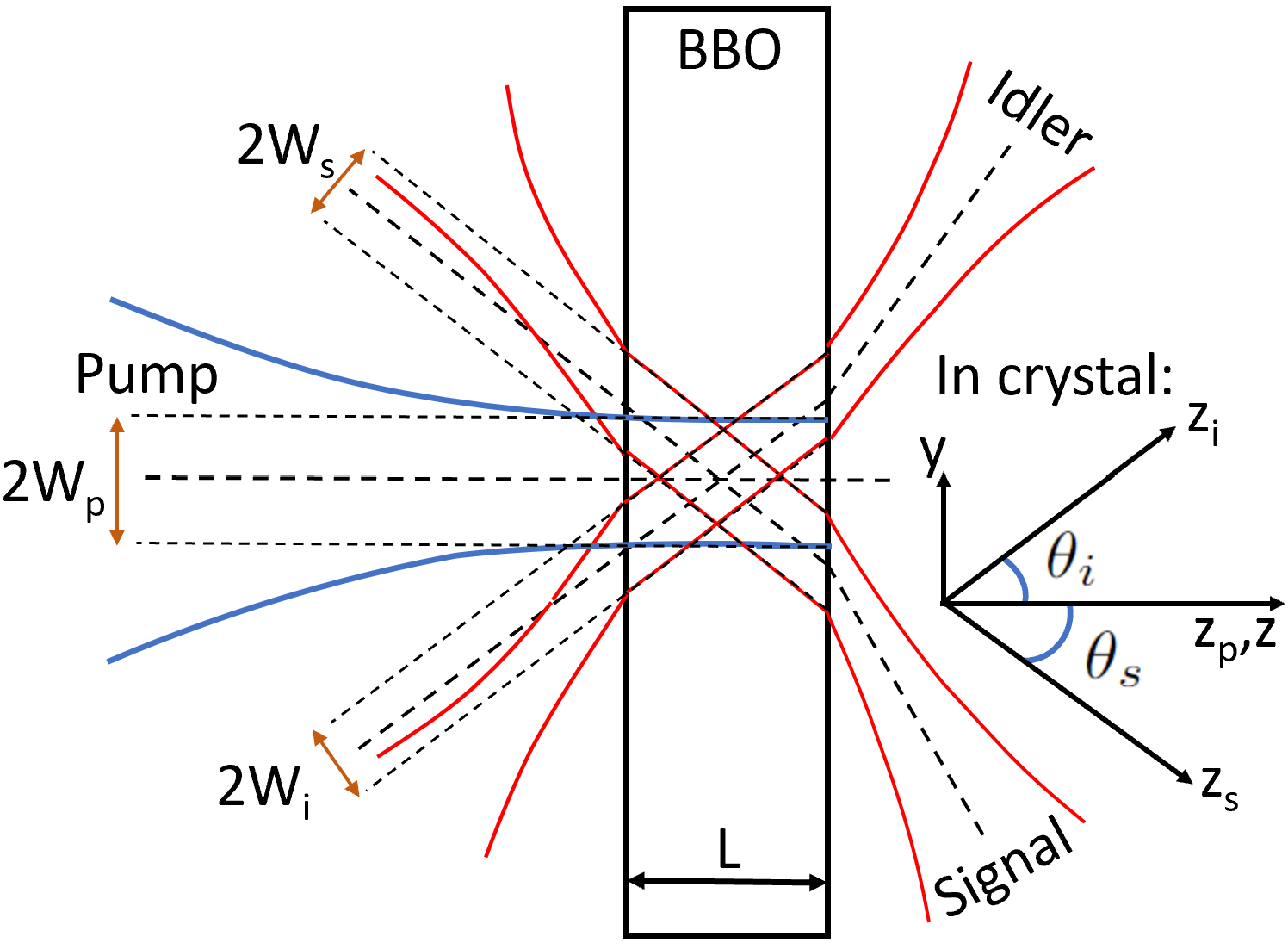}
    \caption{\label{fig:setup} Geometry of the SPDC interaction. The focal conditions of the pump (blue lines indicate $1/e$ field radius of pump), signal, and idler beams (red lines indicate $1/e$ field radius of signal and idler). The beam waist radii are $W_{0,p}$, $W_{0,s}$, $W_{0,i}$. The beams are aligned such that the three modes overlap as much as possible inside the crystal (perfect intersection at the center ($L$/2) of the crystal).}
\end{figure}

The geometry of the fundamental Gaussian modes of the pump beam and the collecting optics are shown in Fig.~\ref{fig:setup}. The emission angle $\theta_s$ ($\theta_i$) for the signal (idler) beam is greatly exaggerated for illustrative purposes. We consider emission angles so small that all modes overlap in the crystal (to within a negligible cross-sectional area mismatch). The crystal cut angle is not shown to keep the figure simple; the crystal axis differs from the $z$-axis by the cut angle for the uniaxial crystal considered here. 

Using $U_j$ to describe the mode of the $j$-field ($j=s$, $i$, and $p$), generally given by Eq.~(\ref{Eq:HGmode}), the joint spectral amplitude is given by \cite{Guilbert},
\begin{multline}
\label{eq:phi1}
    \Phi(\omega_s, \omega_i)= g(\omega_s+\omega_i) \int_{L/2}^{L/2} dz \int_{-\infty}^\infty \int_{-\infty}^\infty dx dy U_p(x,y)U_s(x,y)U_i(x,y)\\ \times \textrm{exp} \left\{ i \left[
    k_pz-k_s(-\sin\theta_sy+\cos\theta_sz)-k_i(-\sin\theta_iy+\cos\theta_iz) \right] \right\},
\end{multline}
The inclusion of the pump spectral distribution $g(\omega_s+\omega_i)$ is where our theory differs from Ling \textit{et al.} and Guilbert \textit{et al.}, where they take $a$ to be a flat distribution. We consider a Gaussian pump distribution
\begin{equation}
\label{Eq:pump_dist}
    g(\omega_s+\omega_i) =\textrm{exp}\left(-\frac{(\omega_s+\omega_i-\omega_{p0})^2}{4B_p^2}\right),
\end{equation}
where $B_p$ is the pump beam spectral bandwidth, $\omega_{p0}$ is the central frequency of the pump. 

The frequency dependence of $\Phi(\omega_i,\omega_s)$ arises from the frequency dependence of $k_j$, given in Eq.~(\ref{eq:wavevector}). The complex exponential in Eq.~(\ref{eq:phi1}) is extracted from Eq.~(\ref{Eq:HGmode}) after our approximations, while the rest of Eq.~(\ref{Eq:HGmode}) remains in each $U_j(x,y)$. Here, $k_j$ for $j = p,\textrm{ }s,\textrm{ and }i$ is the corresponding mode propagation direction.  We rewrite this complex exponential compactly in terms of the phase mismatch $\vec{\Delta k}$. This quantifies momentum conservation for a specific combination of frequencies and indices of refraction.

In the calculation of the pair production rate, we consider all three field modes to be in the fundamental (Gaussian) mode, so that we may write
\begin{multline}
\label{Eq:overlap_int}
    \Phi(\omega_s,\omega_i) =g(\omega_s+\omega_i) \int dz \int dxdye^{i\vec{\Delta k}\textrm{ }\Vec{r}} \textrm{exp}\left(\frac{-(x^2+y^2)}{W_{0,p}^2}\right)\\ \times \textrm{exp}\left(\frac{-(x^2+(y\cos{\theta_s}+z\sin{\theta_s})^2)}{W_{0,s}^2}\right) \textrm{exp}\left(\frac{-(x^2+(y\cos{\theta_i}-z\sin{\theta_i})^2)}{W_{0,i}^2}\right),
\end{multline}
where each term is a Gaussian function in the mode transverse spatial dimension. We consider the fundamental modes because they are most highly occupied when coupling the light into single-mode fiber (which ensures spatial indistinguishability, another important metric for photon pair sources). When we consider signal and idler singles rates, we use the full expression of $U_j(x,y)$ from Eq.~(\ref{Eq:HGmode}). This will allow us to determine the fraction of signal and idler photons that remain within the fundamental mode compared to those that escape to higher-order Hermite-Gauss modes (the heralding efficiency).

In a non-collinear geometry, $\vec{\Delta k}$ is typically split into transverse ($\Delta k_y$) and longitudinal ($\Delta k_z$) components given by
\begin{equation}
\label{Eq:Delta_ky}
    \Delta k_y = k_s \sin \theta_s - k_i \sin \theta_i,
\end{equation}
and
\begin{equation}
\label{Eq:Delta_kz}
    \Delta k_z = k_p - k_s \cos \theta_s - k_i \cos \theta_i.
\end{equation}

The integration over the $x$- and $y$-variables is straightforward and gives us
\begin{equation}
\label{Eq:xyintegrated}
    \Phi(\omega_s,\omega_i)=g(\omega_s,\omega_i)\frac{\pi}{\sqrt{AC}}\textrm{exp}\left(-\frac{\Delta k_y^2}{4C}\right) \int^{L/2}_{-L/2}\textrm{exp}(-H z^2-i\Delta k_z z),
\end{equation}
where
\begin{equation}
    A=\frac{1}{W_{0,p}^2}+\frac{1}{W_{0,s}^2}+\frac{1}{W_{0,i}^2},
\end{equation}

\begin{equation}
\label{Eq:C}
    C=\frac{1}{W_{0,p}^2}+\frac{\cos^2\theta_s}{W_{0,s}^2}+\frac{\cos^2\theta_i}{W_{0,i}^2}.
\end{equation}

\begin{equation}
\label{Eq:D}
    D=\frac{\sin2\theta_s}{W_{0,s}^2}-\frac{\sin2\theta_i}{W_{0,i}^2}
\end{equation}

\begin{equation}
\label{Eq:F}
    F=\frac{\sin^2\theta_s}{W_{0,s}^2}+\frac{\sin^2\theta_i}{W_{0,i}^2}
\end{equation}

\begin{equation}
\label{Eq:H}
    H=F-\frac{D^2}{4C}
\end{equation}
The effect of spatial walk-off (the difference in phase propagation direction and the direction of energy flow) is given in the spatial integral along the $z$-direction in Eq.~(\ref{Eq:xyintegrated}) \cite{Ling}
\begin{equation}
\label{Eq:walk_off}
    \Phi_z=\int^{L/2}_{-L/2}dz\textrm{ exp}(-H z^2-i\Delta k_z z).
\end{equation}

Considering small angles $\theta_s$ and $\theta_i$, and a thin crystal, the walk-off factor ($H$) is approximately zero, and thus we ignore it. This approximation greatly speeds up computation time and only results in a $\sim1\%$ error in $\eta$. The final expression after these approximations is given by
\begin{equation}
\label{Eq:Mode_function}
    \Phi(\omega_s,\omega_i)=\frac{\pi L}{\sqrt{AC}} \textrm{sinc}\left(\Delta k_zL/2\right) \textrm{exp}\left(-\frac{\Delta k_y^2}{4C}-\frac{(\omega_s+\omega_i-\omega_{p0})^2}{4B_p^2}\right).
\end{equation}

Our goal is to express the joint spectral intensity in the form of Eq.~(\ref{Eq:JSIgeneral}). Therefore, we write the components of the mode function in terms of the frequency detuning from the central frequency (the one that yields phase matching at the central pump frequency). Starting with the wave vector magnitudes, we write them as a first-order Taylor expansion about each mode's central frequency as
\begin{equation}
\label{Eq:k_expansion}
    k_j = k_{j0} + N_j \Omega_j,
\end{equation}
where we introduce the inverse group velocity $N_j = (dk_j/d\omega_j)_{\omega_{j0}}$ \cite{Shi}. Here, $\Omega_j = \omega_j - \omega_{j0}$ is the frequency detuning of mode $j$. Using Eq.~(\ref{Eq:k_expansion}) in Eq.~(\ref{Eq:Delta_ky}) and Eq.~(\ref{Eq:Delta_kz}), the zero-order terms drop out and we are left with 
\begin{equation}
\label{Eq:Dky_detuning}
   \Delta k_y = N_s \Omega_s \sin \theta_s - N_i \Omega_i  \sin \theta_i,
\end{equation}
and
\begin{equation}
\label{Eq:Dkz_detuning}
    \Delta k_z = N_p \Omega_p - N_s \Omega_s  \cos \theta_s - N_i \Omega_i  \cos \theta_i.
\end{equation}
We also write the pump spectral distribution from Eq.~(\ref{Eq:pump_dist}) in terms of the signal and idler frequency detunings,
\begin{equation}
\label{Eq:pump_dist2}
    g(\Omega_s,\Omega_i)=\frac{(\Omega_s+\Omega_i)^2}{4B_p^2}.
\end{equation}
Writing everything involved in the mode function in terms of frequency detuning is useful because a frequency uncorrelated JSA (as in Fig.~\ref{fig:JSA}a, a pure source) should be a Gaussian function of signal and idler frequency detuning. The result is 
\begin{equation}
\label{Eq:Mode_function_detuning}
    \Phi(\Omega_s,\Omega_i)=\frac{\pi L}{\sqrt{AC}} \textrm{sinc}\left(\Delta k_zL/2\right) \textrm{exp}\left(-\frac{\Delta k_y^2}{4C}-\frac{(\Omega_s+\Omega_i)^2}{4B_p^2}\right),
\end{equation}
where the phase mismatches are given in Eqs.~(\ref{Eq:Dky_detuning}) and~(\ref{Eq:Dkz_detuning}).

After collecting constants used in previous derivations \cite{Guilbert, Ling}, the emission rate is found using
\begin{equation}
\label{Eq:Joint_rate}
    \mathcal{R} =\frac{\eta_s\eta_iPd_{eff}^2\alpha_s^2\alpha_i^2\alpha_p^2\omega_s\omega_i}{\sqrt{2}\pi^{3/2}\epsilon_0c^3n_sn_in_pB_p} \int_{-\infty}^{\infty}\int_{-\infty}^{\infty}T_s(\omega_s) T_i(\omega_p-\omega_s)|\Phi(\Omega_s,\Omega_i)|^2d\omega_sd\omega_p,
\end{equation}
which gives the pair rate in units of pairs/s. Here, $\eta_j$ are the overall signal/idler path efficiencies, and
\begin{equation}
\alpha_j = \sqrt{\frac{2}{\pi W_{0,j}^2}},
\end{equation}
are the fundamental mode normalization constants [$\alpha^{(0,0)}$ in Eq.~(\ref{Eq:alpha})], $\epsilon_0$ is the vacuum permittivity constant, and the filter transmission functions (which we consider to be ideal unit-efficiency bandpass filters) are $T_j$. Dividing through by the pump power $P$ in milliwatts (mW) and the down-converted spectral filter width in terahertz (THz) gives us the absolute pair emission rate in units of $[\mathcal{R}]=$ pairs/(s$\textrm{ }$mW$\textrm{ }$THz), using the fact that the rate scales linearly with pump power.

In the expression for the bi-photon mode function Eq.~(\ref{Eq:Mode_function}), the factor that involves the beam waists parameter ($C$) is the foundation of our purity maximization technique. By changing the beam waists in $C$, we change the dependence of the spectrum on the transverse phase mismatch. Therefore, the new technique we introduce is only applicable to non-collinear setups. Otherwise, $\Delta k_y = 0$ and the frequency-dependent portion of $\Phi(\Delta k)$ is independent of beam waists. The constants in the form of $1/\sqrt{AC}$ remain, but they do not influence the frequency dependence. In Sec.~\ref{Sec:TuneProp}, we expand this idea. There, we first rewrite the mod-square of Eq.~(\ref{Eq:Mode_function}) into the form of Eq.~(\ref{Eq:JSIgeneral}). This immediately leads to a purity maximization condition ($\delta_{si}\xrightarrow{}0$).

\subsection{Heralding Efficiency}

The heralding efficiency is given by 
\begin{equation}
\label{Eq:heralding_efficiency}
     \eta = \frac{\mathcal{R}}{\sqrt{\mathcal{R}_s \mathcal{R}_i}}, 
\end{equation}
where the $\mathcal{R}_{s(i)}$ is the rate of signal (idler) single-photon production rate and $\mathcal{R}$ is the joint count rate. The spectrum and rate of the singles are derived similarly as the joint spectrum and joint rate above. The difference is that we need to consider the higher-order non-Gaussian modes that the signal (idler) can escape to when calculating the signal (idler) singles spectrum.

We take a summation over all modes $(n,m)$ given in Eq.~(\ref{Eq:HGmode}) and obtain the signal (idler) singles rate
\begin{multline}
\label{Eq:singles_rate}
     \mathcal{R}_{s(i)} = \sum\limits_{n,m=0}^{\infty} \frac{\eta_{s(i)}Pd_{eff}^2(\alpha_{s(i)}^{(n,m)})^2\alpha_{i(s)}^2\alpha_p^2\omega_s\omega_i}{\sqrt{2}\pi^{3/2}\epsilon_0c^3n_sn_in_pB_p} \\ \times \int_{-\infty}^{\infty}\int_{-\infty}^{\infty}T_{s(i)}(\omega_{s(i)})|\Phi_{s(i)}^{(n,m)}(\Omega_s,\Omega_i)|^2 d\omega_{s(i)} d\omega_p.
\end{multline}
In Eq.~(\ref{Eq:singles_rate}), the shape of the pump spectrum is included in $\Phi_{s(i)}^{(n,m)}$, which is the mode function for a given $(n,m)$. From here, $\eta$ is calculated by using both Eq.~(\ref{Eq:Joint_rate}) and Eq.~(\ref{Eq:singles_rate}) in Eq.~(\ref{Eq:heralding_efficiency}). In agreement with \cite{Guilbert}, we find good convergence when we truncate the sum in Eq.~\ref{Eq:singles_rate} for $n$ and $m>10$.

\subsection{State Purity}

The bi-photon state purity is determined by the spectral factorability of the joint spectrum into two separate photon states. In general, the wave function can be expressed using Schmidt decomposition as \cite{Mosley}
\begin{equation}
    \Phi(\omega_s,\omega_i) = \sum\limits_{n}^{}\sqrt{\lambda_n}\phi_n(\omega_s)\psi_n(\omega_i), \hspace{0.3cm} \sum\limits_{n}^{}\lambda_n=1,
\end{equation}
where $\phi_n(\omega_s)$ and $\psi_n(\omega_i)$ are  orthogonal modes in frequency space. Each combination of modes has a weight (Schmidt coefficient) $\lambda_n$, from which we obtain the state purity, 
\begin{equation}
\label{Eq:purity}
    \mathcal{P} = \sum\limits_n^{}\lambda_n^{2},
\end{equation}
which is the inverse of the Schmidt number.

However, Schmidt decomposition is not always analytically possible for the general case; we use the equivalent singular value decomposition method. We calculate a representative set of JSAs for combinations $(\omega_s,\omega_i)$ within the range of our spectral filters. These values correspond to an $n \times m$ matrix of values $\left|\Phi(\omega_{s_n},\omega_{i_m})\right|^2$ which we then perform singular value decomposition. In analogy to Eq.~(\ref{Eq:purity}), the sum of the squared singular values of the joint spectral amplitude matrix is treated as the spectral purity.

\subsection{Symmetrized Fidelity} \label{Sec:Fidelity}
The symmetrized fidelity ($\mathcal{F}$) is a combination of signal and idler fidelities ($\mathcal{F}_s$ and $\mathcal{F}_i$ respectively) to a Gaussian distribution in frequency space. Signal and idler fidelities combine as $\mathcal{F}=\sqrt{\mathcal{F}_s\mathcal{F}_i}$, which, when combined with some simplifications (following Ref.~\cite{Meyer-Scott}), is
\begin{equation}
\label{Eq:Fidelity}
    \mathcal{F}=\eta\left(\frac{2\mathcal{P}}{1+\mathcal{P}}\right).
\end{equation}
From Eq.~\ref{Eq:Fidelity}, $\mathcal{F}$ scales positively with both $\eta$ and $\mathcal{P}$, indicating that maximizing both will approximately result in maximizing $\mathcal{F}$. The goal of high $\mathcal{F}$ clearly illustrates the need to simultaneously have high $\eta$ and $\mathcal{P}$, which is missing from the literature for Type-I SPDC.

\section{Tuning the Properties of Generated Photon Pairs} \label{Sec:TuneProp}
\subsection{Methodology}

Previously, methods of tuning the properties of bi-photon pairs were often difficult to use, and at best inconvenient. Our proposed technique draws from the findings of previous authors and pulls them together. First, we approximate \cite{Shi} 
\begin{equation}
\label{Eq:sinc_exponential}
     \text{sinc}(x)=e^{-(\alpha x)^2},
\end{equation}
where $\alpha=0.455$. This allows us to find an approximate expression for a unit-purity condition. We include corrections to account for inaccuracies of this approximation. Equation~(\ref{Eq:sinc_exponential}) can be used to represent the JSA of the bi-photon (the modulus square of the mode function, Eq.~(\ref{Eq:Mode_function}) above) as
\begin{equation}
\label{Eq:JSA}
    S(\Omega_s,\Omega_i)=\mathcal{N} \textrm{exp}\left(-\frac{\Delta k_y^2}{4C}-\frac{(\alpha\Delta k_zL)^2}{4}-\frac{(\Omega_s+\Omega_i)^2}{2B_p^2}\right).
\end{equation}
The more general form, written in terms of the three frequency detunings, is shown in Eq.~(\ref{Eq:JSIgeneral}). Using the expressions for $\Delta k_y$ and $\Delta k_z$ above [Eq.~(\ref{Eq:Delta_ky}) and Eq.~(\ref{Eq:Delta_kz}), respectively], we have
\begin{equation}
    \delta_s=\frac{N_s^2\text{sin}^2\theta_s}{4C}+\frac{\alpha^2(N_p-N_s\text{cos}\theta_s)^2L^2}{4}+\frac{1}{4B_p^2},
\end{equation}
\begin{equation}
    \delta_i=\frac{N_i^2\text{sin}^2\theta_i}{4C}+\frac{\alpha^2(N_p-N_i\text{cos}\theta_i)^2L^2}{4}+\frac{1}{4B_p^2},
\end{equation}
and
\begin{equation}
\label{Eq:delta_si}
    \delta_{si}=-\frac{\alpha^2(N_p-N_s\text{cos}\theta_s)(N_p-N_i\text{cos}\theta_i)L^2}{2}+\frac{1}{2B_p^2}-\frac{N_s\text{sin}\theta_sN_i\text{sin}\theta_i}{2C}.
\end{equation}

By setting $\delta_{si}=0$, we ensure the signal and idler are separable in frequency and therefore spectrally pure.  When this condition is satisfied, the JSA has an elliptical shape with major and minor axes along $\Omega_s$ and $\Omega_i$ such that no general correlations exist between signal and idler frequencies. In the case when the major and minor axes are equal, the JSA is a two-dimensional Gaussian spectrum, seen in Fig.~\ref{fig:JSA}a. 

When we use $C$ from Eq.~(\ref{Eq:C}) in Eq.~(\ref{Eq:delta_si}) for the non-collinear case (and set $W_{0,i}=W_{0,s}$), the purity condition Eq.~(\ref{Eq:delta_si}) becomes a relationship between Gaussian mode beam waists $W_{0,s}$ and $W_{0,p}$
\begin{equation}
\label{Eq:W_s}
    W_{0,s}= (\text{cos}^2\theta_s+\text{cos}^2\theta_i)^{\frac{1}{2}} \bigg(\frac{N_iN_s\text{sin}\theta_i\text{sin}\theta_s}{\frac{1}{B_p^2}+\alpha^2L^2(N_p-N_i\text{cos}\theta_i)(N_p-N_s\text{cos}\theta_s)}-\frac{1}{W_{0,p}^2}\bigg)^{-\frac{1}{2}}.
\end{equation}
Setting the down-converted collection mode beam waists equal is useful for both maximizing the photon pair rate \cite{Guilbert,Ling}, and for greatly simplifying the form of Eq.~(\ref{Eq:W_s}). Using this approximation, we introduced for $W(z)$ in Eq.~(\ref{Eq:HGmode}) in Sec.~\ref{Sec:FOM}, we rationalize why the two beam waists may be taken as equal (even in the non-degenerate case). The difference in wavelength between the two down-converted modes in the non-degenerate case technically results in a difference in spot size evolution, according to our definition of $W_j(z)$. We consider this negligible for large $z_{r,j}$ and a thin crystal, just as we assume $W_j(z)\approx W_{0,j}$ in the crystal.

The purity condition Eq.~(\ref{Eq:W_s}), which results from requiring $\delta_{si}=0$, can only be achieved for certain combinations of values. For example, at $\theta_j=0$ (collinear setups), the term involving $C$ is zero, so the beam waists are no longer a part of the purity condition. Equation~(\ref{Eq:delta_si}) then yields a fixed relationship between $B_p$ and $L$ (where they are inversely proportional). Although it is still possible to reach high $\mathcal{P}$ using this relationship, it lacks the additional degree-of-freedom from the beam waists. Thus, better combinations of $\eta$, $\mathcal{P}$, and $\mathcal{R}$ are sacrificed when using the collinear case.  See, for example, Bennick's work \cite{Bennink}, where he studies how the Gaussian beam waists effects $\mathcal{P}$ and $\eta$.

We first iteratively maximize our figures-of-merit in $B_p$, $L$, cut angle, filter widths, etc. Then, we tune the beam waists, which (as shown in \cite{Vicent}) should allow $\mathcal{P}=1$. Furthermore, we push to attain the highest possible combinations of $\eta$, $\mathcal{P}$, and $\mathcal{R}$.

To achieve this goal, we first iterate over values of $W_{0,p}$ to maximize $\mathcal{R}$. We use this value of $W_{0,p}$ in the purity condition [Eq.~(\ref{Eq:W_s})] to find the corresponding $W_{0,s}$. Then, we detune $W_{0,s}$ and find the value of $W_{0,s}$ that results in the highest $\mathcal{P}$. Note that Eq.~(\ref{Eq:W_s}) tends to overestimate the ideal value of $W_{0,s}$. These steps are repeated, re-optimizing the auxiliary variables (not beam waists) to optimize overall performance. There are many trade-offs between the figures-of-merit. Many such relationships will be explored in Sec.~\ref{Sec:Generalize}. 

As an example, we use this optimization method for the case of a $400\textrm{ }-\mu m$-long BBO crystal, a crystal cut angle detuning (angular perturbation from the cut angle that produces collinear signal and idler beams) of $2.2^\circ$, flat-top unit efficiency spectral filters on the down-converted beams with a half-width of $2.5 \textrm{ THz}$ ($\Delta\omega_s=\Delta\omega_i=5 \textrm{ THz}$ illustrated in Fig.~\ref{fig:DC_filters}), a spectral filter on the pump beam twice as wide as that used on the down-converted beams; and a $8 \textrm{-THz}$-bandwidth pump beam.

The configuration that led to the best optimization initially included a much wider pump bandwidth (by a factor of $\sim4$). However, to maximize the efficiency of down-conversion, there should be minimal temporal walk-off between the pump and down-converted wave packets. The group velocities are different enough between pump and down-converted photons that the pump needed to be temporally broadened, and therefore spectrally narrowed. The $8 \textrm{-THz}$-bandwidth pump beam that we use gives good overlap between wave packets at the end of their travel through the crystal, and therefore minimizes temporal walk-off.

\subsection{Non-collinear Degenerate.}

Non-collinear setups are achieved by altering the crystal's cut angle. Collinear setups require a specific cut angle unless non-critical phase matching is used (crystal axis and pump propagation direction are orthogonal). For a non-collinear geometry, we consider a small $2.2^{\circ}$ detuning from the collinear cut angle (determined iteratively as discussed in Sec.~\ref{Sec:TuneProp}). Our calculations assume a nearly complete transverse spatial overlap between the pump and down-converted beam collection modes inside the crystal, which is still valid for our choice of cut angle and crystal length. Others \cite{Guilbert} have considered setups that have approximately a $\approx3^\circ$ full angle between signal and idler outside the crystal, but we see no reason to sacrifice $\mathcal{P}$ through limiting transverse phase mismatch when pump beam spatial walk-off is affected in such a minor way.

\begin{figure}[ht]
    \centering
    \includegraphics[width = .47\textwidth]{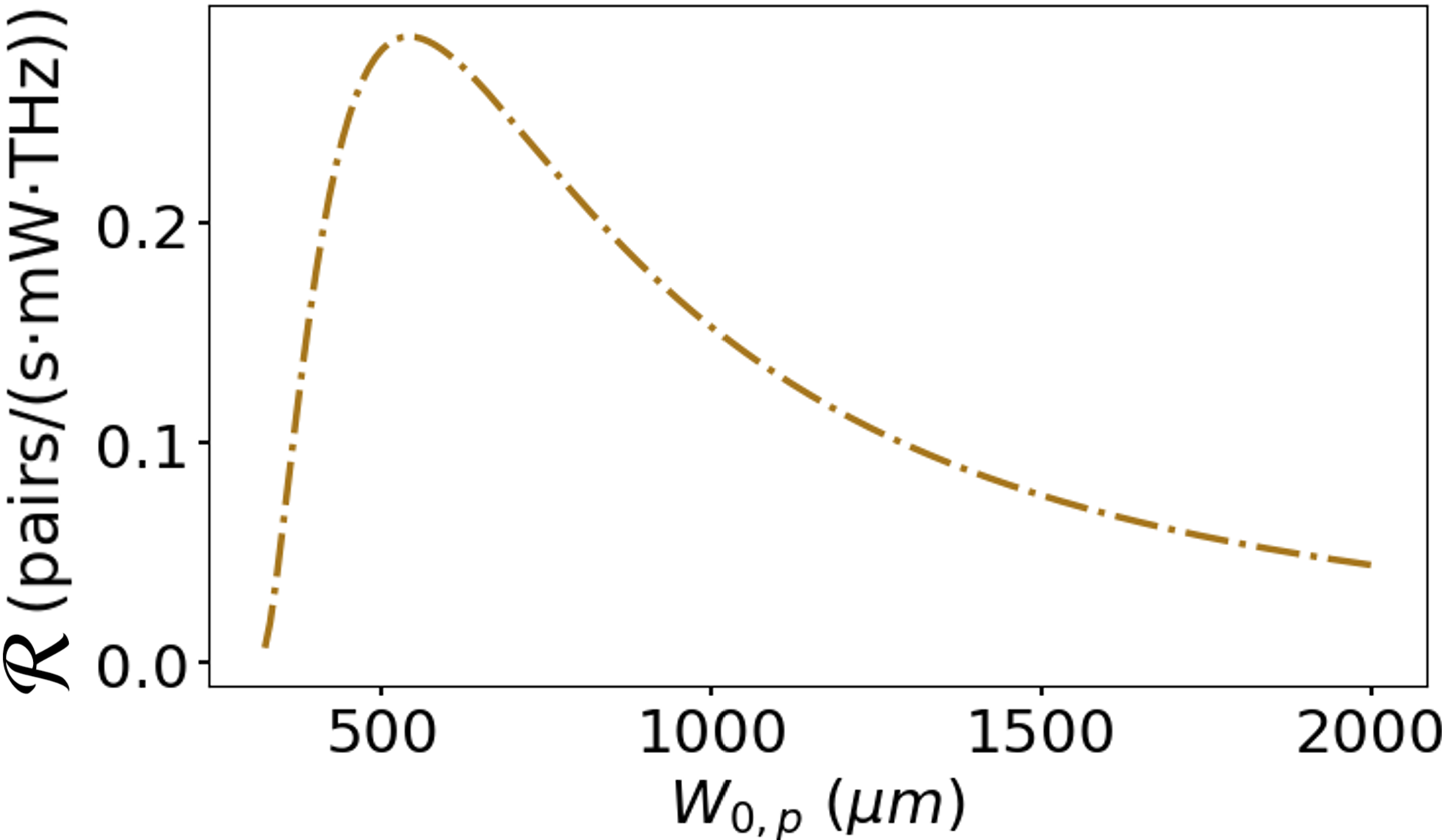}
    \caption{\label{fig:RvsW0} The pair production rate as a function of the pump beam waist.}
\end{figure}

\begin{figure}[ht]
    \centering
    \includegraphics[width = .47\textwidth]{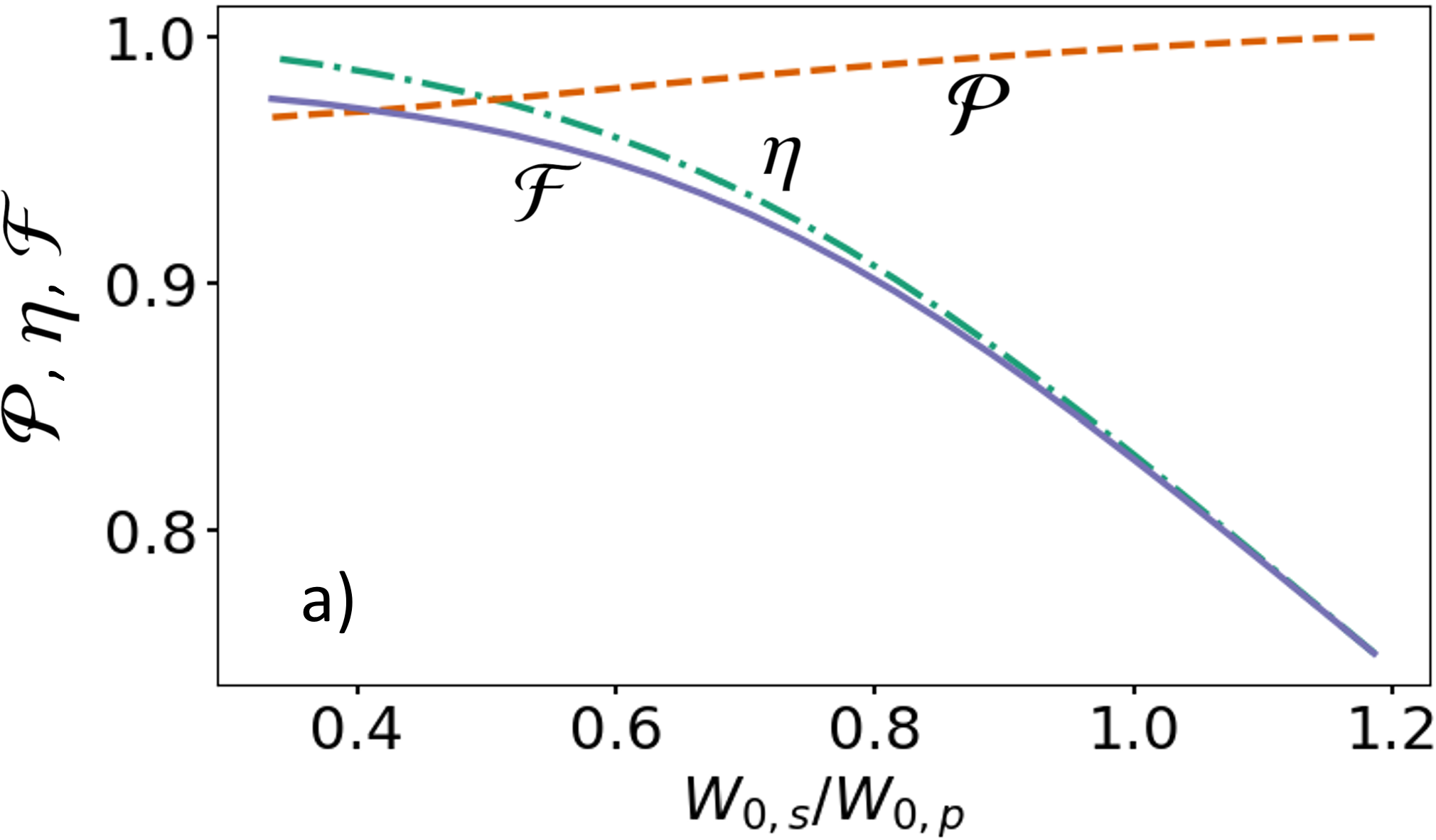}
    \hfill
    \includegraphics[width=.47\textwidth]{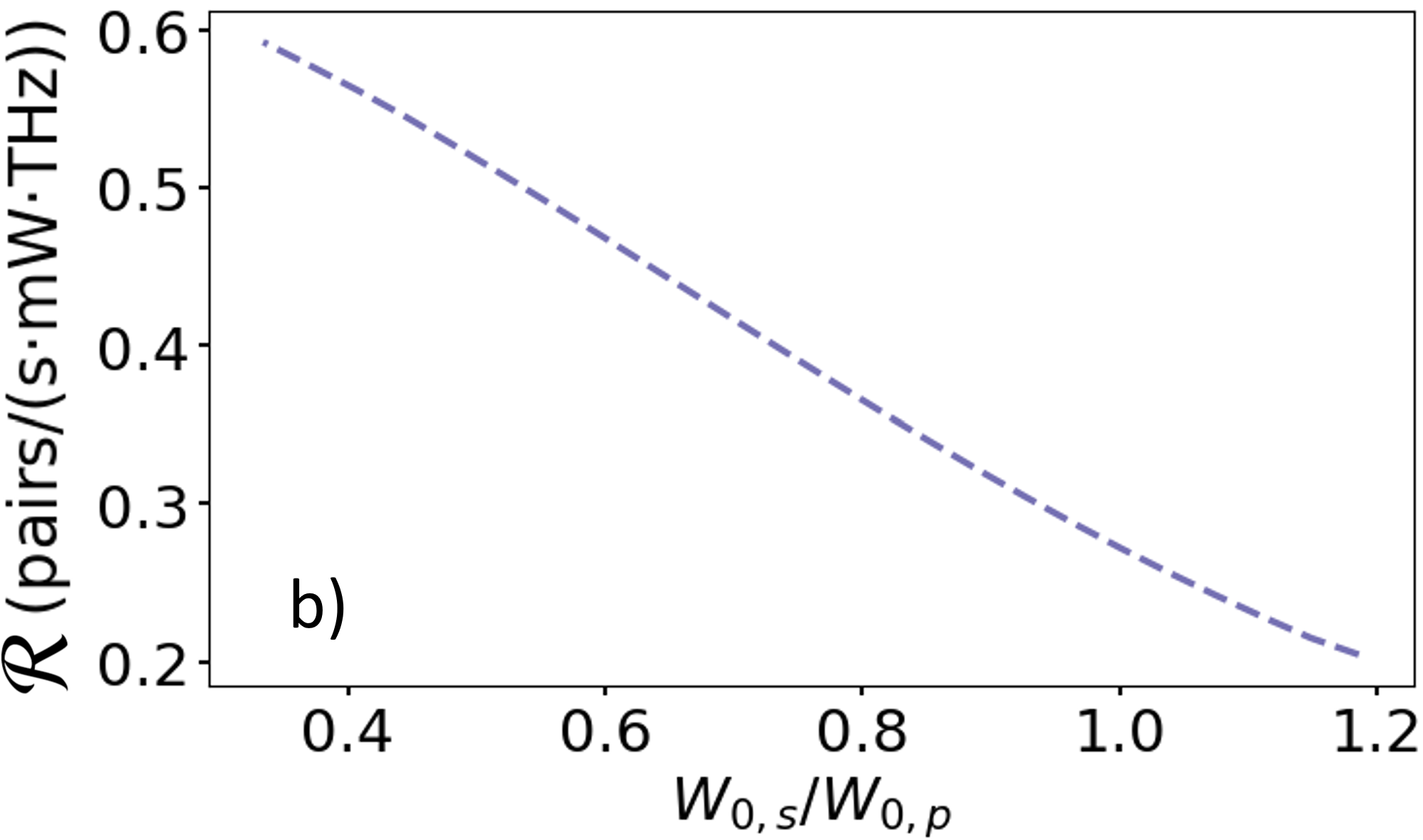}
    
    \caption{\label{fig:noncol_rate} Non-Collinear Degenerate. a) Heralding efficiency, biphoton state purity, and symmetrized fidelity as a function of $W_{0,s}/W_{0,p}$. b) Pair production rate as a function of $W_{0,s}/W_{0,p}$.}
\end{figure}

In the degenerate case, we assume the signal and idler pairs are created at a wavelength centered at 710 nm. As shown in Fig.~\ref{fig:RvsW0}, the maximum $\mathcal{R}$ occurs at a pump beam waist of $W_{0,p}=540\textrm{ }\mu m$. Using the purity maximizing condition Eq.~(\ref{Eq:W_s}), we find $W_{0,s}$ given this $W_{0,p}$. Then, we detune $W_{0,s}$ to predict how the figures-of-merit change as a function of $W_{0,s}$ in Fig.~\ref{fig:noncol_rate} (where the horizontal axis the ratio of $W_{0,s}$ to the ideal rate-maximizing $W_{0,p}$). In our setup, Eq.~(\ref{Eq:W_s}) predicted unit $\mathcal{P}$ at $W_{0,s}=540.5\textrm{ }\mu$m. This is (to within $1\textrm{ }\mu$m) equal to $W_{0,p}$, so that $W_{0,s}/W_{0,p} = 1$. At this point, $\mathcal{P}=0.995$. While $\mathcal{P}$ is very nearly $1$, it continues to grow for looser collection beam focusing ($W_{0,s}/W_{0,p} > 1$).

In Fig.~\ref{fig:noncol_rate}a we see that the intersection of $\mathcal{P}=\eta=0.97$ occurs at $W_{0,s}/W_{0,p}=0.52$, corresponding to $W_{0,s}=280.8\textrm{ }\mu$m for our setup with $W_{0,p}=540.\textrm{ }\mu$m. Altering the setup to achieve $\mathcal{P}\approx1$ only requires adjusting $W_{0,s}/W_{0,p}\approx1.2$, or $W_{0,s}$ to $648\textrm{ }\mu$m. Likewise, near-unit $\eta$ can be achieved for tighter focusing of down-converted collecting optics. 

\subsection{Non-collinear Non-degenerate.}

\begin{figure}[ht]
    \centering
    \includegraphics[width = .47\textwidth]{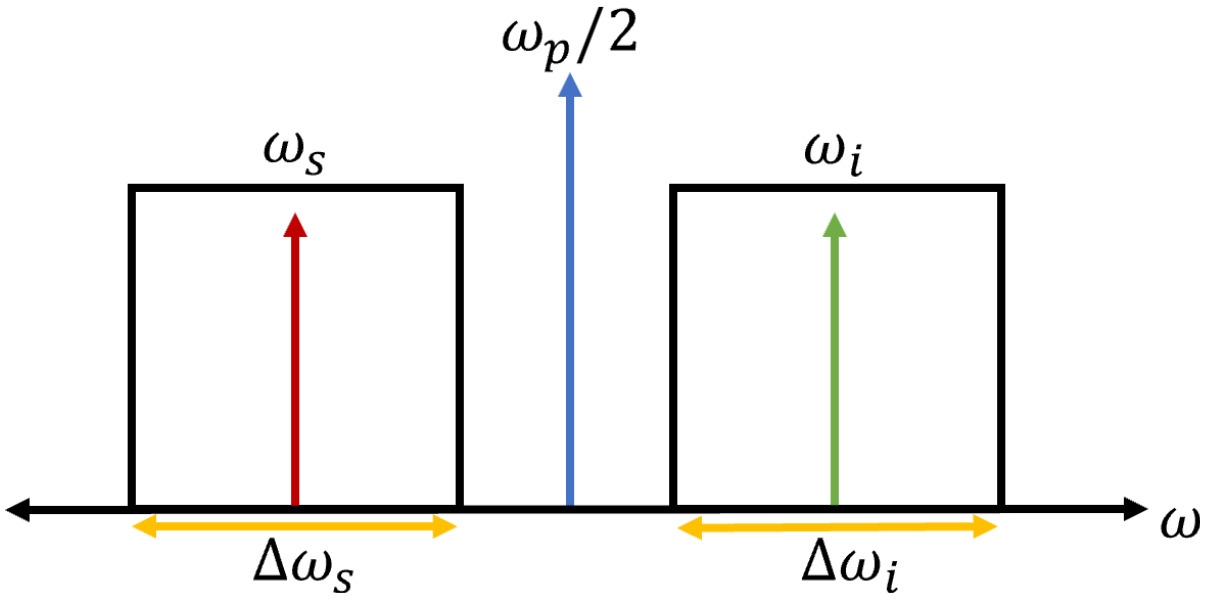}
    \caption{\label{fig:DC_filters} The unit efficiency bandpass filters considered for the non-degenerate case. They are spaced evenly about the degenerate frequency ($\omega_p$) and $\Delta \omega_{s(i)}=10 \textrm{ THz}$.}
\end{figure}

\begin{figure}[ht]
    \centering
    \includegraphics[width = .47\textwidth]{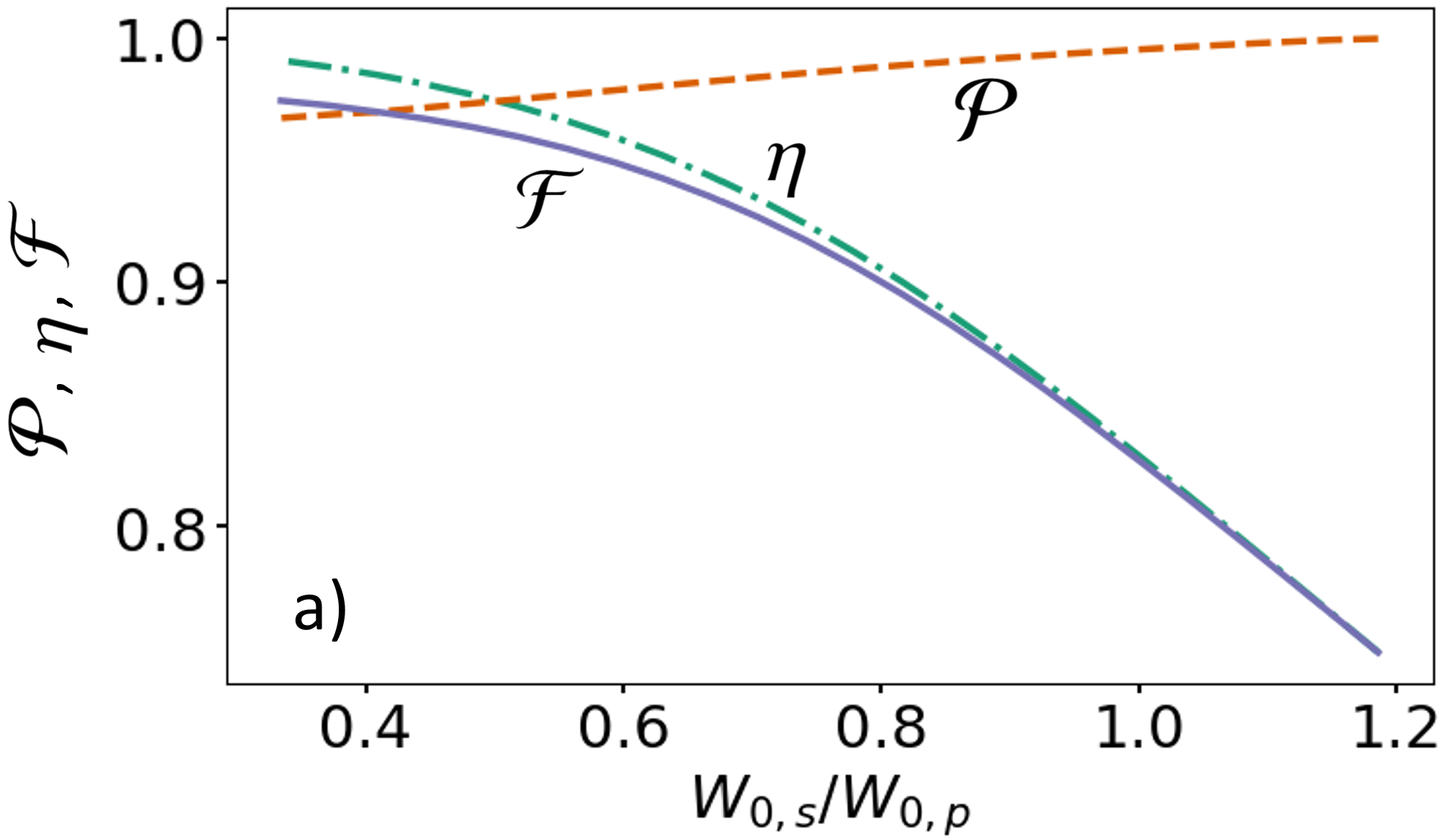}
    \hfill
    \hfill
    \includegraphics[width=.47\textwidth]{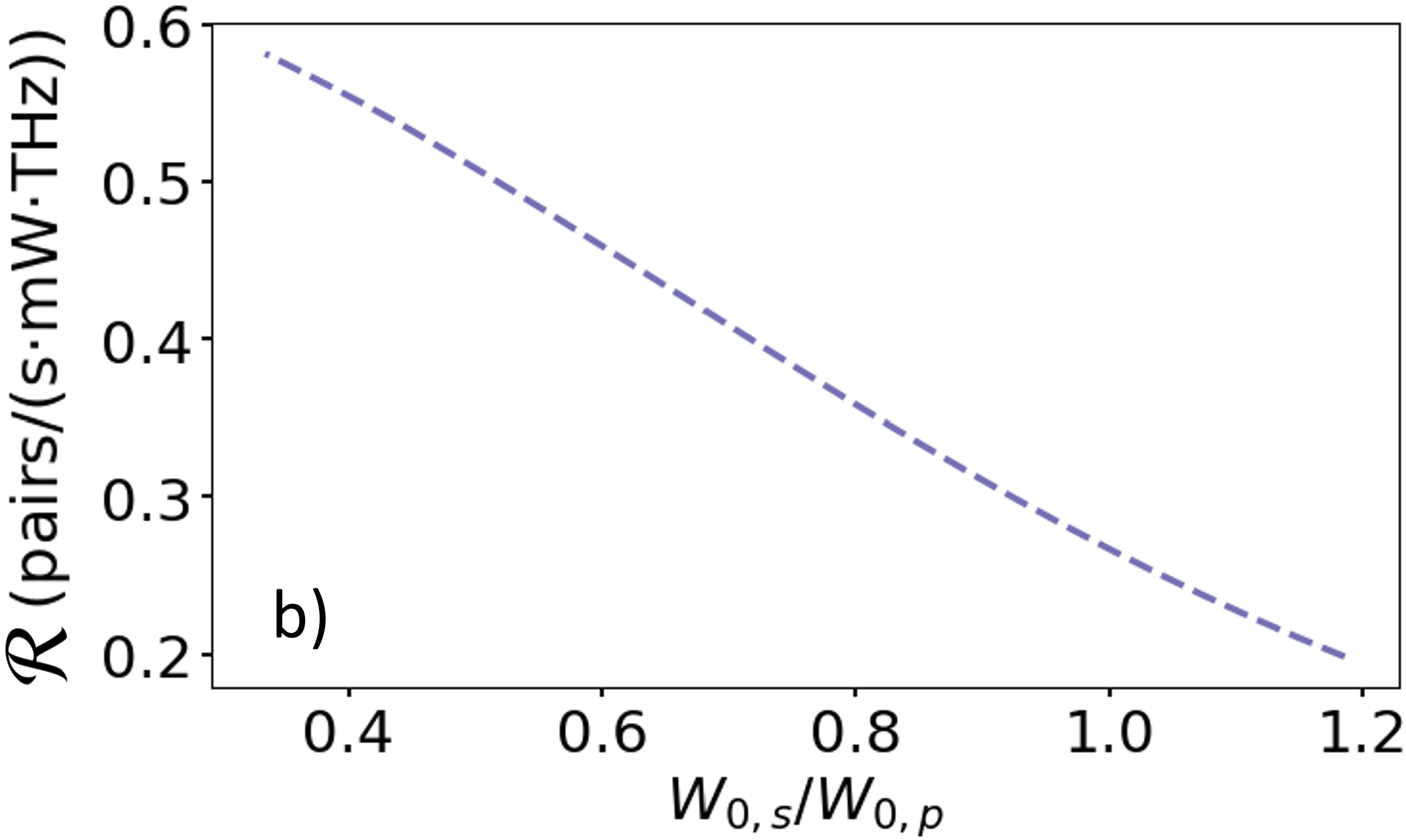}
    \caption{\label{fig:noncol_nondeg_rate} Non-Collinear Non-Degenerate. a) Heralding efficiency, biphoton spectral purity, and symmetrized fidelity as a function of $W_{0,s}/W_{0,p}$. b) Pair production rate as a function of $W_{0,s}/W_{0,p}$.}
\end{figure}

A non-degenerate SPDC source is also sometimes referred to as ``bichromatic.'' To keep $\eta$ as high as possible, the filters are equidistant on either side of half the central pump frequency (energy conservation dictates down-converted pair symmetry about the degenerate frequency), as shown in Fig.~\ref{fig:DC_filters}. These bands of light sometimes overlap and can be separated using spectral filtering, but this is unnecessary here due to the non-collinear geometry. The wavelength and bandwidth of the pump remain the same as in our degenerate setup. We chose the central signal wavelength to be 850 nm and the corresponding energy-conserving central wavelength of the idler is 609.6 nm. As in the degenerate case, $W_{0,p}\approx 540\textrm{ } \mu m$ maximizes $\mathcal{R}$ (the plot looks nearly identical to Fig.~\ref{fig:RvsW0}, so we do not include it here). Using the purity condition Eq.~(\ref{Eq:W_s}) to calculate $W_{0,s}$ given $W_{0,p}$, we look at the relationships between $W_{0,s}/W_{0,p}$ (where we use $W_{0,p}=540\textrm{ }\mu$m) and our figures-of-merit in Fig.~\ref{fig:noncol_nondeg_rate}. 

Equation~(\ref{Eq:W_s}) is based on approximation in Eq.~(\ref{Eq:sinc_exponential}), so it does not guarantee unit $\mathcal{P}$. Similar to the degenerate case, it predicts unit $\mathcal{P}$ at $W_{0,s}/W_{0,p}\approx1$, but our calculations show $\mathcal{P}=0.995$ at this point. Purity continues to grow slowly as $W_{0,s}/W_{0,p}$ increases. At the largest $W_{0,s}/W_{0,p}$ that we plot (1.2, corresponding to $W_{0,s}\approx648\textrm{ }\mu\textrm{m}$), our model predicts $\mathcal{P}=0.9997$. This is approximately $20\%$ off of what Eq.~(\ref{Eq:W_s}) predicts. Using Eq.~(\ref{Eq:sinc_exponential}) is therefore not practical for obtaining exact results; however, it is a useful tool to get close to unit $\mathcal{P}$. To illustrate the trade-offs in the pursuit of perfect $\mathcal{P}$, for $\mathcal{P}=0.9997$ we have $\eta=0.750$, while for $\mathcal{P}=0.990$ we have $\eta=0.901$.

\section{Generalization to other variables} \label{Sec:Generalize}

Although we mainly focused on the relationship between the down-converted/pump beam waists and the figures-of-merit, there are trade-offs to explore in adjusting the constants appearing in Eq.~(\ref{Eq:Joint_rate}),~(\ref{Eq:singles_rate}) and~(\ref{Eq:delta_si}). The most straightforward way to understand their effect on $\mathcal{P}$ is to think of how well $\delta_{si}=0$ can be enforced for a given set of variables. Parameters that are in the singles and pair production rate [Eq.~(\ref{Eq:singles_rate}) and~(\ref{Eq:Joint_rate}), respectively] affect $\eta$, as seen in Eq.~(\ref{Eq:heralding_efficiency}). Naturally, the parameters changing in the equation for $\mathcal{R}$ Eq.~(\ref{Eq:Joint_rate}) will affect it.

Increasing the down-converted frequency filter widths while keeping the pump filter bandwidth twice as wide as the down-converted filter widths decreases $\mathcal{P}$ and $\eta$ while increasing $\mathcal{R}$. The primary purpose of the down-converted beam filters is to eliminate any remaining frequency correlations in the region of large $\Omega_s$ and $\Omega_i$, increasing $\mathcal{P}$. However, strong filtering (decreasing filter widths) significantly lowers $\mathcal{R}$, as mentioned in Sec.~\ref{sec:Background}.

As the cut angle detuning increases (decreases), the angle between the two down-converted beams increases (decreases). The cut angle must be large enough to introduce a transverse phase mismatch that can be tuned by adjusting the beam waists, using our purity condition Eq.~(\ref{Eq:W_s}). At small values of $\theta_j$ (for $j=s,i$), Eq.~(\ref{Eq:W_s}) can only be satisfied for large pump spectral bandwidths.

Looking beyond experimental logistics, increasing the pump spectral bandwidth decreases $\mathcal{R}$. A larger cut angle will introduce a more substantial Poynting vector walk-off. This will make our approximation of negligible walk-off between Eq.~(\ref{Eq:walk_off}) and Eq.~(\ref{Eq:Mode_function}) inaccurate. Generally, greater walk-off will negatively impact $\eta$ and $\mathcal{R}$ due to less beam overlap inside the crystal (thus, the issue also worsens with greater crystal length).

In tuning the cut angle, it is also important to remember that $d_{eff}$, the refractive indices, and the phase mismatch vary. We considered a cut angle detuning of $2.2^{\circ}$, which results in the walk-off angle of the pump beam only growing by $0.111^{\circ}$ (or $2.63\%$ of the collinear pump walk-off). We find that including the walk-off term results in a small ($<1\%$) difference in the overall joint count rate. This indicates that, for our chosen value of crystal cut angle, the spatial walk-off has a negligible effect on the system. In contrast, a larger cut angle or a longer crystal will experience a more noticeable negative effect.

When increasing the crystal length, we continually adjust the optimal down-converted beam waists $W_{0,s}$ to maintain high purity. We write the JSA in terms of its length-dependent terms as 
\begin{equation}
\label{Eq:JSA_vs_L}
    S(L)=|\Phi(L)|^2= \mathcal{N} \frac{L^2}{AC} \textrm{sinc}^2\left(\Delta k_zL/2\right) \textrm{exp}\left(-\frac{\Delta k_y^2}{2C}\right),
\end{equation}
where $\mathcal{N}$ includes length-independent constants. Next, we consider perfect phase matching (at the center frequency of the ideal filters, $\Delta k_z = 0$) and drop the constants, yielding 
\begin{align}
\label{Eq:JSA_vs_L_approx1}
    S(L)\approx \frac{L^2}{AC} \textrm{exp}\left(\frac{-(\Delta k_y L)^2}{2}\right).
\end{align}
Looking at the (approximate) purity maximization condition [Eq.~(\ref{Eq:W_s})], $W_{0,s}$ scales approximately linearly with $L$. The coefficients $A$ and $C$ scale as $1/W_{0,s}^2$, and thus as $1/L^2$. Therefore, we finally approximate the JSA as
\begin{align}
\label{Eq:JSA_vs_L_approx2}
    S(L)\approx L^6 \textrm{exp}\left(\frac{-(\Delta k_y L)^2}{2}\right).
\end{align}
These approximations make it clear that, for increasing $L$ starting at zero, $\mathcal{R}$ [which scales as $S(L)$ with length] rises and then falls after the inflection point where the exponential overtakes the polynomial factor. This goes against what we typically expect for a phase-matched interaction \cite{boyd}, $\mathcal{R} \sim L^2$. The heralding efficiency also decreases with greater crystal length, which is also due to the larger $W_{0,s}$ required for a longer crystal (in agreement with \cite{Guilbert}). A longer crystal also means less beam spatial overlap in the crystal, eventually making our approximation of Eq.~(\ref{Eq:walk_off}) for negligible walk-off inaccurate. 

Bulk BBO approaches that primarily aim to maximize count rates have achieved nearly $1,000\textrm{ pairs}/(\textrm{s} \textrm{ mW})$ using a $2-\textrm{mm}$-long crystal \cite{Kurtsiefer}. Our approach, which produces count rates of $\sim 1\textrm{ pairs}/(\textrm{s} \textrm{ mW} \textrm{ THz})$, can still be useful for many single-photon level experiments where photon qualities are more important than quantities. The damage threshold for BBO is on the order of $\textrm{GW}/\textrm{cm}^2$ \cite{Bhar}, so our approach can be employed in a lab setting to produce many thousands of high-quality entangled photons per second for a properly chosen laser (rates which are easily perceivable when using good quality single photon detectors). One of the limiting factors of $\mathcal{R}$ is the crystal's length; we use a crystal a few times shorter than most other setups due to the negative effects on $\eta$ and $\mathcal{R}$ from increasing crystal length. Another limiting factor is down-converted filter bandwidth\textemdash while we do not filter strongly, it still limits count rates.

The trade-offs in altering the physical parameters of the system make it convenient to choose values for each that approximately optimize the system and then tune the Gaussian beam waists for fine control over the figures-of-merit. This is the technique we used in our simulation to produce Figs.~\ref{fig:RvsW0},~\ref{fig:noncol_rate} and~\ref{fig:noncol_nondeg_rate}. Changing the pump, signal, and idler wavelengths has a minor effect on the figures-of-merit, which can be counteracted by changing beam widths. It is much easier to alter beam widths than it is to change other physical parameters such as the length of the crystal, thus supporting our claim that the method presented here provides high spectral flexibility.

\section{Conclusions} \label{Sec:Conclusion}

We show that changing beam focal parameters makes it possible to achieve high heralding efficiency and purity in a Type-I non-collinear SPDC setup. This technique will enable researchers to produce a pure single-photon source using basic optical equipment and a bulk nonlinear crystal, once initial parameters such as crystal length are chosen. Our model enables us to predict that, for both non-collinear degenerate and non-degenerate setups at unit purity, the heralding efficiency can be as high as $\approx0.90$, and purity and heralding efficiency intersect at $\approx0.97$. Our setup is therefore a cost-effective way to bypass more expensive or complicated setups with theoretically comparable results. We remove the need to specially engineer crystals and waveguides for purity or heralding efficiency, simplifying the process of attaining high purity and efficiency. To the best of our knowledge, this is the best spectral purity and heralding efficiency predicted to date in Type-I SPDC. Our method also eliminates the need to strongly filter the down-converted beams. Additionally, our setup can operate over a broad range of pump wavelengths, unlike crystal dispersion-engineered pure single photon sources.

\begin{backmatter}
\bmsection{Acknowledgement}
The authors thank Dr. Gregory Lafyatis for fruitful discussions that improved this manuscript's clarity and overall quality.

\bmsection{Disclosures}
The authors declare no conflicts of interest.

\bmsection{Data availability}
The date and code used to produce the figures in this paper are available from the authors on reasonable request. 
\end{backmatter}

\bibliography{bibliography}
\end{document}